\documentclass[fleqn,usenatbib]{mnras}

\usepackage{newtxtext,newtxmath}

\usepackage[T1]{fontenc}

\DeclareRobustCommand{\VAN}[3]{#2}
\let\VANthebibliography\thebibliography
\def\thebibliography{\DeclareRobustCommand{\VAN}[3]{##3}\VANthebibliography}

\usepackage{graphicx}	
\usepackage{amsmath}	
\usepackage{booktabs}   
\usepackage{subcaption}  
\usepackage{xcolor}

\newcommand{\Rf}{\text{Ra}_\text{F}}

\title[Rotating IH\&C Convection]{Heat Transport and Dissipation in 2.5D Rotating Internally Heated and Cooled Convection}

\author[T. Joshi-Hartley et al.]{
Tom Joshi-Hartley,$^{1}$\thanks{E-mail: tj294@exeter.ac.uk}
Matthew K. Browning,$^{1}$
Laura K. Currie,$^{2}$
Neil T. Lewis,$^{1}$
Benjamin P. Brown,$^{3}$\newauthor
\ and Simon R. W. Lance$^{1}$ 
\\
$^{1}$Physics and Astronomy Department, University of Exeter, Stocker Rd, Exeter EX4 4QL, UK\\
$^{2}$Department of Mathematical Sciences, Durham University, Durham DH1 3LE, UK\\
$^{3}$Astrophysical and Planetary Sciences, University of Colorado at Boulder, Boulder, United States \\
}

\date{Accepted XXX. Received YYY; in original form ZZZ}

\pubyear{\the\year{}}

\begin{document}
\label{firstpage}
\pagerange{\pageref{firstpage}--\pageref{lastpage}}
\maketitle

\begin{abstract}
Models of astrophysical convection, such as mixing length theory, typically assume that the heat transport is independent of microphysical diffusivities. Such `diffusion-free' behaviour is, however, not observed in numerical simulations employing standard fixed-flux or fixed-temperature boundary conditions, except possibly in extreme parameter regimes that are computationally expensive to achieve. Recent numerical and experimental work has suggested that internally heated and cooled convection can exhibit diffusion-free scalings in more numerically accessible regimes. Here, we present direct numerical simulations of 2.5D Cartesian rotating thermal convection driven by an internal heating and cooling function. The use of distributed heating and cooling functions alleviates sharp thermal boundary layers that would otherwise be present, allowing the flows to be simulated with modest computational resources.  We show that for high Rossby numbers this set-up recovers mixing length theory scalings for the heat transport. The velocity amplitudes, in contrast, are observed to display diffusion-limited scalings. By comparing against boundary driven rotating convection, we show that internally heated cases have a larger fraction of their thermal dissipation occuring in the bulk of the fluid. We suggest this is connected to the increased convective efficiency observed in these cases. Our results indicate that 2.5D internally heated convection can be used as a computationally inexpensive test-bed to investigate some aspects of diffusion-free heat transport.
\end{abstract}




\section{Introduction}

All stars are convective in some part of their interiors at some point in their lifetimes \citep[e.g.][]{Kippenhahn13}. As such, an understanding of astrophysical convection is essential when constructing models of stars. 
However, our understanding of stellar and solar convection is still incomplete. For example, in the so-called `radius inflation problem', low-mass stars are observed to have larger radii than current stellar evolution models predict; both rotation and magnetism have been proposed as mechanisms to suppress convection and lead to radius inflation \citep[e.g.][]{Chabrier07, Kraus11, Ireland18}, but the requisite field strengths and/or rotation rates are still under debate. Additionally, although many features of solar convection can be captured by simulations \citep{Nordlund09, Kapyla23}, deep models of the Sun do not (for example) reliably recover solar-like differential rotation (fast equator, slow pole) at the solar rotation rate, and do over-estimate the deep convective velocities \citep{Lord14}. These discrepancies can be mitigated, with methods including increasing the rotation rate \citep[e.g.][]{Augustson15} or decreasing the convective luminosity \citep{Hotta15}, but they speak to the gaps in our knowledge of astrophysical convection.

Convection in models of stellar evolution is often parameterised by mixing length theory \citep{Gough76, Joyce23}, and as such the convection is `diffusion-free' \citep[e.g.][]{Kraichnan62, Spiegel63, Spiegel71} - that is, the molecular diffusivities are so low that they are not a limiting factor in the efficiency of the convection \citep[][]{Lepot18}. There is a keen interest in finding numerical set-ups which allow for diffusion-free convection, as the molecular diffusivities in stars are expected to be low enough that heat transport would be diffusion-free \citep{Ludwig99}, allowing for extrapolation to astrophysical regimes and helping constrain our understanding of complicating effects. Common numerical set-ups however, such as Rayleigh-Bénard Convection (RBC) - where a temperature gradient imposed across the fluid drives the convection - do not exhibit diffusion-free behaviour. Instead, thermal boundary layers are observed to form which throttle the efficiency of the heat transport \citep{Malkus54, Belmonte93, Grossmann00, Li12}.

Recently, both lab experiments \citep[e.g.][]{Lepot18} and numerical simulations \citep[e.g.][]{Kazemi22} have managed to achieve non-rotating diffusion-free heat transport scaling through the use of internal heating and cooling (IH\&C). Additionally, prior works focusing on rotation \citep[e.g.][]{Stellmach14, Plumley16, Song24a} have found behaviour that tends towards the diffusion-free regime, including in 3D spherical shells \citep{Gastine16}. Some astrophysically-motivated papers have simulated rotating convection driven through internal heating and cooling functions \citep[e.g.][]{Barker14, Currie20}, but only recently has this set-up been investigated experimentally \citep{Bouillaut21}, and been complemented with direct numerical simulations (DNS) \citep{Hadjerci24}.

In this paper, we explore the effects of the internal heating and cooling function from \citet{Kazemi22} in rotating 2.5D Cartesian simulations at low spatial resolutions, allowing for a wide parameter sweep. We investigate a range of properties of these simulations to understand how well this computationally inexpensive set-up recovers the predictions of diffusion-free convection. In particular, we investigate heat transport scalings, velocity amplitudes, and dissipation, and find good agreement between the heat transport and the predictions of MLT, suggesting that some of the behaviour observed in 3D simulations can be recovered in simpler 2.5D set-ups. By studying both no-slip and stress-free boundaries, we also clarify the effect of boundary conditions on obtaining diffusion-free behaviour.

In \hyperref[sec:Terminology]{Section \ref{sec:Terminology}} we describe our governing equations, and provide some context onto expected heat transport scalings in different regimes, and summarising experimental and numerical evidence for these. In \hyperref[sec:Problem]{Section \ref{sec:Problem}} we outline our computational set up, and introduce our parameter space. \hyperref[sec:Results]{Section \ref{sec:Results}} contains our results, while \hyperref[sec:Discussion]{Section \ref{sec:Discussion}} outlines our conclusions.

\section{Terminology \& Theoretical Expectations}
\label{sec:Terminology}
Here we briefly outline control and diagnostic parameters we are using, and how they have been explored in prior works.

\subsection{Governing Equations}
\label{sec:2.1}

We consider a horizontally periodic Cartesian box, with a rotation vector $\mathbf{\Omega} = \Omega_0 (0, \sin\theta, \cos\theta)$, where $\Omega_0$ is the rotation rate and $\theta$ is the co-latitude, so $\theta=0$ corresponds to alignment of the gravity and rotation vectors.  We employ the Boussinesq approximation, and adapt the set-up used in \citet{Kazemi22}, non-dimensionalising our equations with the viscous diffusion time $\tau_\nu = d^2 / \nu$, where $d$ is the vertical length scale (height of the box) and $\nu$ is the kinematic viscosity, and the pressure scale $\rho_0 \nu^2 / d^2$, where $\rho_0$ is the reference density. This results in the dimensionless equations for velocity $\mathbf{u}(\mathbf{x}, t)$, pressure $P(\mathbf{x}, t)$ and temperature $T(\mathbf{x}, t)$
\begin{equation}
\label{eq:MassConservation}
    \mathbf{\nabla} \cdot \mathbf{u} = 0,
\end{equation}
\begin{equation}
\label{eq:MomentumEq}
    \partial_t \mathbf{u} + (\mathbf{u}\cdot\mathbf{\nabla})\mathbf{u} = -\nabla P + \frac{\Rf}{\text{Pr}} T \mathbf{\hat{z}} + \nabla^2\mathbf{u} - \text{Ta}^{\frac{1}{2}}\mathbf{\Omega}\times\mathbf{u},
\end{equation}
\begin{equation}
\label{eq:TempEvolution}
    \partial_tT + (\mathbf{u}\cdot \nabla)T = \frac{1}{\text{Pr}}\nabla^2T + H(z),
\end{equation}
where $z$ is the vertical component of the spatial co-ordinate $\mathbf{x}$, $H(z)$ is the vertical heating profile with dimensions of $Q$, the Prandtl number $\text{Pr} = \nu/\kappa$, where $\kappa$ is the thermal diffusivity, which is fixed to $\text{Pr}=1$ for this work, and the Taylor number $\text{Ta} = 4\Omega_0^2d^4/\nu^2$ gives the balance between rotational and viscous forces, such that a higher Taylor number corresponds to stronger rotational effects. The Taylor number is related to the Ekman number, another commonly used control parameter for rotating convection, by $\text{Ek} \equiv \text{Ta}^{-\frac{1}{2}}$.

The other dimensionless control parameter, $\Rf$, is a flux-based Rayleigh number where $\Rf = \frac{\alpha g_0 d^5 Q}{\nu^2 \kappa}$, with $\alpha$ the coefficient of thermal expansion, $g_0$ the magnitude of gravitational acceleration in the $\mathbf{-\hat{z}}$ direction, and $Q$ a heating rate, with units of temperature per time. Similarly to \citet{Kazemi22}, this differs from the RBC Rayleigh number $\text{Ra}= (\alpha g_0 d^3 \Delta T) / (\kappa \nu)$ because the temperature scale $T_\text{C}=d^2Q/\nu$ rather than the dimensional temperature difference between boundaries $\Delta T$. This means Ra is dynamically determined during the simulation, and can be extracted as a diagnostic quantity.

To characterise the vertical heat transport, we use the Nusselt number $\text{Nu}$, which is a measure of how much heat is carried by convection rather than conduction. We define 
\begin{equation}
    \text{Nu}= 1 + \frac{\langle F_\text{conv} \rangle_V}{\langle F_\text{cond} \rangle_V},
\end{equation}
where $\langle \cdot \rangle_V$ denotes a volume and time average, the convective flux $F_\text{conv} = \langle w T\rangle_V$, where $w$ is the vertical velocity component, and the conductive flux $F_\text{cond} = \langle - \frac{\partial T}{\partial z} \rangle_V$. The output Rayleigh number is related to the Nusselt number through $\text{Ra} \propto \Rf / \text{Nu}$, where the constant of proportionality does not change between simulations.

To characterise the degree of rotational constraint, we use the Rossby number Ro, such that
\begin{equation}
    \label{eq:Rossby}
    \text{Ro} = \langle \frac{|\nabla \times \mathbf{u}|}{2|\mathbf{\Omega}|}\rangle_V,
\end{equation}
where a value of $\text{Ro}<1$ corresponds to rotationally-dominated flow, and $\text{Ro} > 1$ corresponds to convectively dominated flow. This definition of the Rossby number scales like $u/l$, where $u$ and $l$ are velocity and length scales respectively. For this reason, this Rossby number predominantly characterises the effects of rotation on small scales.

Another commonly used control parameter is the convective Rossby number $\text{Ro}_\text{c} = \sqrt{\text{Ra}/\text{Pr}\text{Ta}}$, which is often used to parameterise the rotational contribution to the flow dynamics \citep[e.g.][]{Gilman77, Julien12, Ecke23}. In our work, where Ra is in output parameter, the similar quantity $\text{Ro}_\text{c, F} = \sqrt{\Rf/\text{Pr}{\text{Ta}}}$ would be our input convective flux-based Rossby number. As outlined in \cite{Anders19}, Ro is a dynamically calculated \textit{output} parameter and better captures the `true' rotational contribution to the flow than the \textit{input} parameter $\text{Ro}_{\text{c, F}}$, which only provides an estimate of how rotationally constrained the flow is likely to be.

Many previous investigations into diffusion-free convection construct modified diffusivity-free variants of the Rayleigh and Nusselt number, $\text{Ra}*$ and $\text{Nu}*$ \citep[e.g.][]{christensen_zonal_2002,aubert_steady_2005,christensen_scaling_2006,Bouillaut21,Hadjerci24}. These parameters are created by combining the standard non-dimensional numbers to remove the dependence on thermal or viscous diffusion. In a regime where viscosity does not play an important role the flow properties should be observed to scale like $\text{Ra}*$ rather than $\text{Ra}$. These diffusivity-free parameters can then also be combined into a diffusivity-free Reynolds number and Rossby number, as in \citet{kapyla_convective_2024}. In this work, we investigate the diffusivity-free flux based Rayleigh number $\Rf* = \Rf\text{Ek}^3/\text{Pr}$ and the diffusivity-free Nusselt number $\text{Nu}* = \text{NuEk}/\text{Pr}$.

\subsection{Expected scalings}
\label{sec:2.2}
It is expected for any Boussinesq convecting system that the heat transport $\text{Nu}$ should scale with the buoyancy driving $\text{Ra}$ according to some exponent, $\text{Nu} \propto \text{Ra}^\gamma$ for suitable definitions of Nu \citep{Goluskin15}. The value of $\gamma$ in different set-ups is an ongoing area of study, and some of the primary relevant findings are outlined below.

In typical RBC, assuming that convection causes the bulk of the fluid to be isothermal \citep[e.g.][]{Malkus54}, all of the vertical temperature gradient $\Delta T$ is held in thermal boundary layers of width $\lambda$, in which all of the heat is transported by conduction. In this set-up, $\text{Nu}=(Fd)/(\kappa \Delta T) = d/\lambda$, as $F = \kappa \Delta T/\lambda$ is the flux carried in the boundary layer. The Rayleigh number contained within the boundary layer scales with the width of the boundary layer $\lambda$ rather than $d$, such that $\text{Ra}^\text{BL} = (\alpha g_0 \lambda^3 \Delta T) / (\kappa \nu)$, and $\text{Ra} / \text{Ra}^\text{BL} = d^3 / \lambda^3 = \text{Nu}^3$. By assuming that $\text{Ra}^{BL}$ is of order $\text{Ra}_c$, the known critical Rayleigh number corresponding to the onset of convection, it follows that $\text{Nu} \propto \text{Ra}^\frac{1}{3}$. Since $\text{Ra}\propto\Rf / \text{Nu}$, in our system this scaling in our system:
\begin{equation}
    \label{eq:DLNR_Nu}
    \text{Nu} \propto \Rf^\frac{1}{4}.    
\end{equation}
As summarised in \citet{Grossmann00} \citep[see also][]{Siggia94}, many simulations and experiments of RBC have found values of $\gamma$ close to 1/3, varying between $0.28 \lesssim \gamma \lesssim 0.31$. We will refer to this as the boundary-limited scaling.

When taking into account the effects of rotation, we will expect the Nusselt number to scale $\sim \text{Ra}^\gamma \text{Ta}^\beta$, rather than just $\sim \text{Ra}^\gamma$. One model, as suggested by \citet{King12}, assumes that the thermal boundary layer is stabilised by rotation, and that the temperature drop across the fluid is still predominantly in the boundary layers. Within this stable boundary layer the heat transport occurs primarily by conduction, hence the condition for stability in the boundary layers implies that $\text{Nu}\propto \text{Ra}^3 \text{Ek}^4$. In terms of $\Rf$ and Ta, this is equivalent to
\begin{equation}
    \label{eq:DLR_Nu}
    \text{Nu} \propto \Rf^\frac{3}{4} \text{Ta}^{-\frac{1}{2}},
\end{equation}
giving us a diffusion-limited scaling for rotating convection, which we will refer to as the King scaling. \citet{King12} found good agreement between this model and their experiments and simulations of rotating convection.

In contrast to these boundary-limited scalings, many authors have considered the possibility of diffusion-free behaviour in both rotating and non-rotating convection. In the non-rotating case, it has been argued that with sufficient horizontal velocity shear caused by turbulence in the bulk (i.e., at high $\text{Ra}$), the boundary layers outlined in \citet{Malkus54} could themselves become turbulent, increasing the efficiency of the heat transport and resulting in $\gamma=0.5$ \citep[e.g.][]{Kraichnan62}. This is sometimes referred to as the `ultimate' regime in the physics community, where the total heat transport through the domain is independent of the viscosity. Obtaining this regime in numerical simulations has been the focus of many studies, outlined in \autoref{sec:Priors}. This also matches the predictions of mixing length theory (MLT) \citep[][]{Bohm-Vitense58, Spiegel63}, a commonly used parameterisation of convection used in many stellar models, so we will refer to it as the `mixing-length' regime. With our flux-based Rayleigh number, this diffusion-free non-rotating scaling is represented as
\begin{equation}
    \label{eq:DFNR_Nu}
    \text{Nu} \propto \Rf^\frac{1}{3}.
\end{equation}

In the diffusion-free rotating regime, the Nusselt number scaling can be estimated analytically \citep[e.g.,][]{Julien12, Aurnou20}. By supposing $\text{Nu} \propto (\text{Ra} / \text{Ra}_c)^\gamma$, using the prediction from linear theory that $\text{Ra}_c \propto \text{Ek}^\frac{-4}{3}$, and assuming that total heat flux is independent of diffusivities, then $\text{Nu} \propto \text{Ra}^\frac{3}{2} \text{Ek}^2$. In our set-up, this is
\begin{equation}
    \label{eq:DFR_Nu}
    \text{Nu} \propto \Rf^\frac{3}{5} \text{Ta}^{-\frac{2}{5}},
\end{equation}
which represents a diffusion-free rotating scaling relation, matching the predictions of rotating mixing length theory \citep[RMLT, e.g.,][]{Stevenson79}. In these models the temperature drop across the fluid occurs predominantly in the bulk, rather than being throttled by boundary layers.

An overview of the $\text{Nu}$ scalings presented here, in terms of both $\Rf$ and $\text{Ra}$, can be found in \autoref{tab:ScaleTable}.

\begin{table}
    \centering
    \caption{The different Nusselt number scaling laws referred to throughout this work, both in terms of the Rayleigh number $\text{Ra}$ and the flux-based Rayleigh number $\Rf$.}
    \resizebox{\columnwidth}{!}{%
    \begin{tabular}{|c|c|c|c|c|}
    \hline
         Description & $\text{Ra}$ & $\Rf$ & Diffusion-Free & Rotationally \\ 
                     &       &       &                & Constrained \\ \hline 
         Boundary-Limited & $\text{Nu} \propto \text{Ra}^\frac{1}{3}$ & $\text{Nu} \propto {\Rf}^\frac{1}{4}$ & No & No\\
         King & $\text{Nu} \propto \text{Ra}^3 \text{Ek}^4$ & $\text{Nu} \propto {\Rf}^\frac{3}{4} \text{Ta}^\frac{-1}{2}$ & No & Yes\\
         MLT & $\text{Nu} \propto \text{Ra}^\frac{1}{2}$ & $\text{Nu} \propto {\Rf}^\frac{1}{3}$ & Yes & No\\
         RMLT & $\text{Nu} \propto \text{Ra}^\frac{3}{2} \text{Ek}^2 $ & $\text{Nu}\propto {\Rf}^\frac{3}{5}\text{Ta}^\frac{-2}{5} $ & Yes & Yes\\ \hline
    \end{tabular}
    }
    \label{tab:ScaleTable}
\end{table}

\subsection{Scalings Observed in Experiments and Simulations}
\label{sec:Priors}

The predictions of the boundary-limited, non-rotating scaling given in \autoref{eq:DLNR_Nu}, have been well tested. In some previous experimental works \citep[e.g.][]{Castaing89, Chilla93, Glazier99}, scaling exponents have been found around $\alpha={2}/{7}$, and the $\alpha=0.3$ scaling has been obtained both numerically \citep[e.g][]{Amati05} and experimentally \citep[e.g.][]{Rossby69, Niemela06, Funfschilling05, Sun05, He12, Cheng15}.

The effects of rotation have been studied in some detail, experimentally \citep[e.g.][]{Rossby69, Cheng15} and numerically \citep[e.g.][]{King12, Stellmach14, Plumley16, Song24a}. As reviewed by \citet[][see their Fig. 2]{Plumley19}, rotating convection tends to display both rotationally-constrained and convectively-dominated regimes. 
Drawing on a wide array of simulations and experiments, they argued that the convectively-dominated regime follows the boundary-limited scaling in \autoref{eq:DLNR_Nu}, while the rotationally-constrained regime depends on the boundary conditions - no-slip flows scale in a diffusion-limited manner like \autoref{eq:DLR_Nu} and free-slip flows scale in a diffusion-free manner, as in \autoref{eq:DFR_Nu}. Note, however, that the no-slip spherical shell simulations of \citet{Gastine16} recovered the diffusion-free RMLT prediction for high Taylor numbers and high supercriticalities. Likewise, recent simulations in \cite{Song24a, Song24b, Song24c}, which have achieved Rayleigh numbers up to $5\times10^{13}$ and Taylor numbers to $4\times10^{16}$, find that for the most extreme parameters the diffusion-free regime is achieved with no-slip boundary conditions.

 High Rayleigh number experiments \citep[e.g.][]{Ahlers09, He12} have suggested that sufficiently turbulent non-rotating convection will eventually transition to the diffusion-free regime. However, obtaining $\text{Ra} \gtrsim 10^{12}$ in 3D computational simulations is difficult, and in experimental works requires specific conditions and media \citep{Niemela00}. To overcome this limitation, some works aim to alter the system to achieve the `mixing-length' regime by avoiding the formation of thin thermal boundary layers \cite[e.g.][]{Barker14}. The present work primarily builds on the findings of \citet{Kazemi22}, who reported that through using a `net-zero' heating and cooling function (the amount of heating matches the amount of cooling, such that the net flux injected into the domain is zero), diffusion-free scaling can be obtained in non-rotating, 3D simulations. Experimentally, internal heating has also been investigated in lab experiments, where light sensitive dye is mixed into the fluid and heated via a strong light-source \citep[e.g.][]{Lepot18, Bouillaut19, Bouillaut22}. In these lab-experiments, this radiative heating has been observed to drive convection that exhibits diffusion-free scalings. However, works that look at constant internal heating \citep[e.g.][]{Goluskin12, Goluskin16, Powers23}, find boundary-limited scaling laws. 

 The interplay between rotation and internal heating and cooling has also been investigated both numerically and, more recently, experimentally. 3D numerical simulations display diffusion-free scaling both at polar latitudes ($\theta=0^\circ$) \citep{Barker14} and non-polar latitudes \citep{Currie20} ($\theta \neq 0^\circ$). \cite{Bouillaut21} presented the first experimental set-up of radiatively heated rotating convection, and obtained diffusion-free results. \citet{Hadjerci24} compared these lab experiments of radiatively driven convection to 3D DNS, and found diffusion-free scalings in both the rotationally constrained and convectively-dominated regimes.

In summary, prior work has suggested that diffusion-free scalings can be obtained by convection in some regimes, but the conditions under which this occurs are not yet clear. Typically such behaviour has only been realised at high supercriticalities and in 3D, which is difficult to achieve. The role of momentum boundary conditions also remains unclear. These issues motivate our work below.

\section{Computational Set-up}
\label{sec:Problem}

We consider \hyperref[eq:MassConservation]{Equations 1, 2, \& 3} with internal heating and cooling and rotation. To avoid the computational cost associated with expanding our domain into three dimensions, we simulate in 2.5 dimensions - a 2D box in $y$ and $z$ is simulated, but the velocity $\mathbf{u}=u\mathbf{\hat{x}} + v\mathbf{\hat{y}} + w\mathbf{\hat{z}}$ retains its $x$-component, and is a function of $t$, $y$ and $z$ only. The 2D box has an aspect ratio of $\Gamma$, and is periodic in $y$. We perform simulations both with the no-slip boundary conditions $\mathbf{u}=0$ at the top and bottom boundaries, as in \citet{Kazemi22}, and also with free-slip conditions, where $u_z = 0$ and $\partial_z u_x = \partial_z u_y = 0$, as they are more astrophysically relevant \citep[e.g.][]{Yadav13}. We enforce an insulating boundary condition at $z=0$, so $\partial_z T=0$, and fix $T=0$ at the top boundary $z=1$. To avoid horizontal streaming suppressing the convection with our free-slip boundary conditions, we fix the aspect ratio $\Gamma=4$, as in \citet{Fuentes21}.

The heating term referred to on the right-hand side of \autoref{eq:TempEvolution} is the net-zero heating and cooling profile from \citet{Kazemi22}
\begin{equation}
    \label{eq:HeatEq}
    H(z) = a e^{-\frac{z}{\ell}} - 1,    
\end{equation}
where $a=1/[\ell(1-e^{-1/\ell})]$ so that $\langle a e^{-z/\ell} \rangle = 1$. The exponential length scale, $\ell$, is fixed such that $\ell=0.10$. This profile gives a heating zone at the bottom of the box and a cooling zone elsewhere. 

\begin{figure}
    \centering
    \includegraphics[width=\columnwidth]{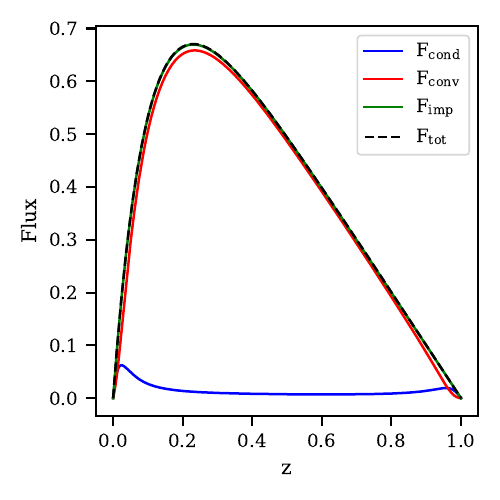}
    \caption{Time- and velocity-averaged flux profiles for a free-slip case with $\text{Ta}=10^9$ and $\Rf=9.1\times10^8$ plotted against height. The imposed flux from the H\&C function is shown with the green line, and the total flux is shown with the dashed black line, which converges to the imposed flux. The blue line shows the conductive flux, which is close to zero in the bulk but rises at each boundary. The red line gives the convective flux.}
    \label{fig:fluxes}
\end{figure}

We used the pseudo-spectral code Dedalus \citep{Burns20} with the $3/2$ dealiasing rule to perform direct numerical simulations of this set-up with Taylor numbers of $[10^6, 10^7, 10^8, 10^9]$ for a range of $\Rf$. Cases were tested with $N_y$ and $N_z$ set to $128\times64$, and the resolution increased if the simulation was not fully resolved. To ensure time-averages were computed over a long enough period to capture dynamics, simulations were run for a minimum of 2.5 viscous times, and the flux balances were investigated. In this set-up, a constant total flux throughout the domain is not expected. Instead the total flux imposed by the heating and cooling ($F_\text{imp}(z') = \int_0^{z'} H(z) dz$) was calculated, and compared with the total flux ($F_\text{cond} + F_\text{conv}$), as shown in \autoref{fig:fluxes}. The time-average was deemed to be sufficient when $\frac{|F_\text{imp} - F_\text{tot}|}{F_\text{tot}} < 2.5 \% $, and the simulations satisfied the spatio-temporal power integrals discussed in \hyperref[sec:Dissipation]{Section \ref{sec:Dissipation}}. A full overview of cases is given in \hyperref[app:SimParams]{Appendix A}.

\section{Results}
\label{sec:Results}

\subsection{Regimes of Convection}
\begin{figure}
    \centering
    \includegraphics[width=\columnwidth]{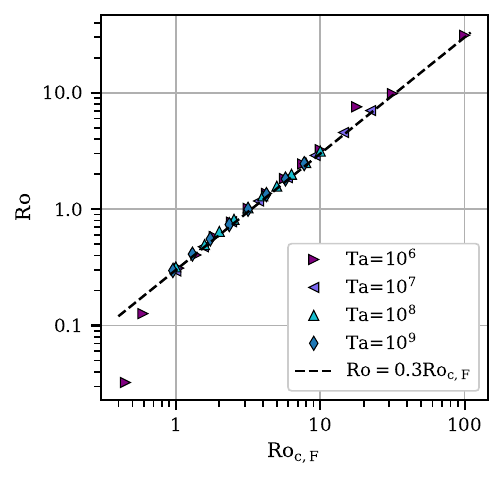}
    \caption{The correlation between the input convective flux-based Rossby number $\text{Ro}_\text{c, F}$, and the output Rossby number Ro.}
    \label{fig:RoRoc}
\end{figure}

The simulations here all rotate, but in some regimes rotation plays little dynamical role, whereas in others it dominates the dynamics.
To determine which cases were rotationally-dominated and which were convectively-dominated, we used the output Rossby number Ro defined in \autoref{eq:Rossby}. In \autoref{fig:RoRoc}, we compare Ro to the convective flux-based Rossby number defined relative to input parameters, where we find that although the two parameters are well correlated, $\text{Ro}_\text{c, F}$ over-estimates Ro (under-estimates degree of rotational constraint) by roughly a factor of three. Throughout our Results and Discussion, when we refer to a ``high" or ``low" Rossby number, we will be referring to the dynamically determined Rossby number $\text{Ro}$.

\begin{figure*}
    \centering
    \includegraphics[width=\textwidth]{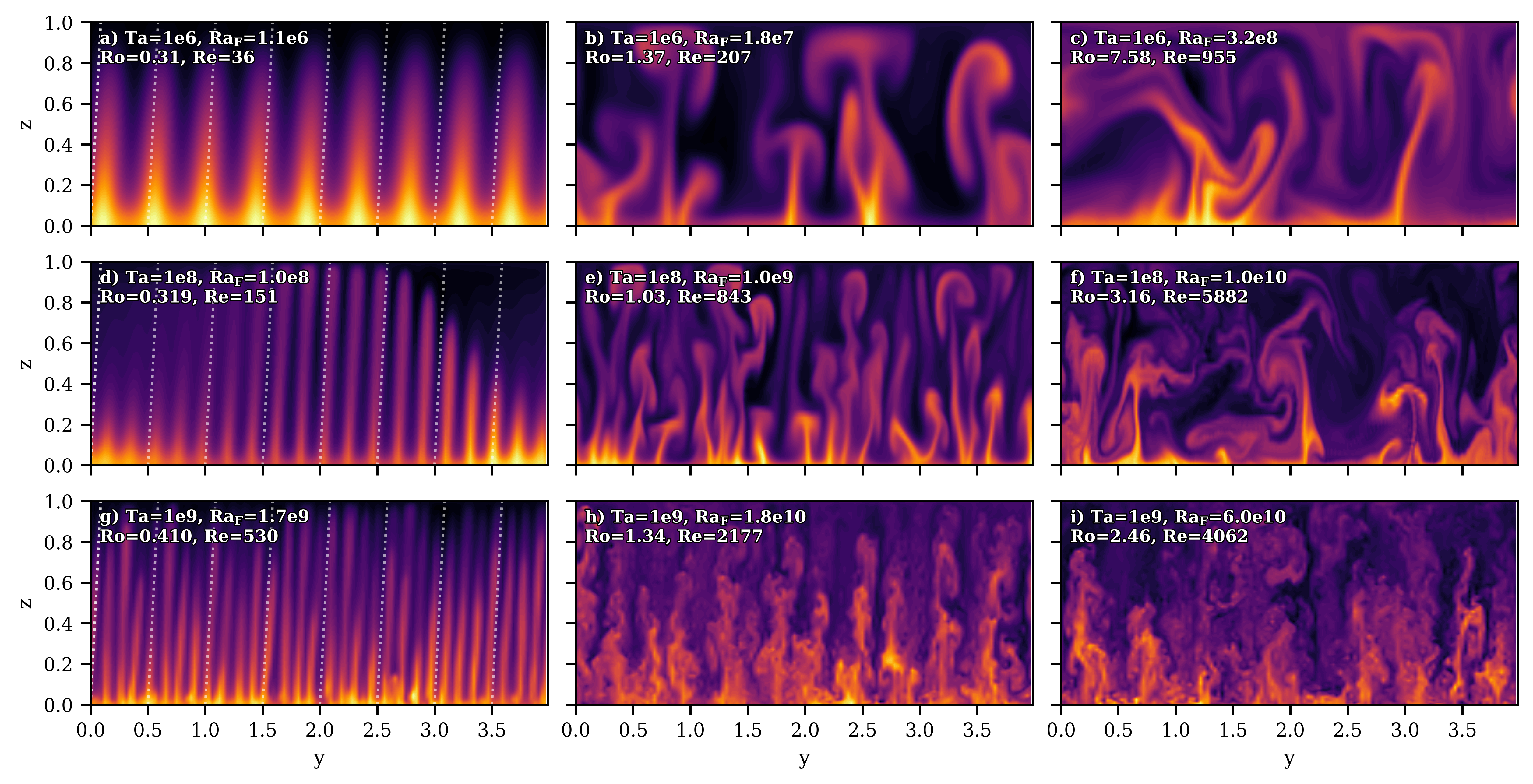}
    \caption{Snapshots of the temperature field of a range of simulations for selected $\Rf$ and Ta, with free-slip boundary conditions. The Rossby number and Reynolds number are shown for each case. It can be seen that for $\text{Ro}<1$, the flow aligns itself into convective fingers, aligned with the rotation axis at 5$^\circ$ to the vertical (shown by white dashed lines on the left-most figures). At $\text{Ro}>1$, the flow shows turbulent convective plumes. For cases at $\text{Ro}\simeq1$, close to the `knee' in \autoref{fig:FS_NuRa}, the flow is in between the two, with semi-turbulent convective fingers. This behaviour is seen for both free-slip and no-slip boundaries.}
    \label{fig:DynamicCompar}
\end{figure*}

Observing the temperature field displays a stark difference in the dynamics between rotationally-dominated and convectively-dominated regimes. \autoref{fig:DynamicCompar} shows the temperature field for representative cases with free-slip boundary conditions. Low Ro cases display hot convective fingers, aligned with the axis of rotation at $5^\circ$ to the vertical (as shown by the white dashed lines), and the horizontal length scale of these fingers decreases as Taylor number increases - i.e. the faster the box is rotating, the more convective fingers can be seen in the box, implying a higher wavenumber dominant mode of convection. The convectively-dominated (high Ro) cases bear more resemblance to standard boundary driven RBC, showing turbulent hot plumes of convection rising from the bottom of the box.

We also use the kinetic energy power spectra to investigate the dynamics of the system, and to confirm that our simulations are adequately resolved. To calculate power spectra we first output the velocity fields at the mid-plane of the box ($z=0.5$) and take the Fourier transform of the velocity, defined as
\begin{equation}
    \mathbf{U}(k_y) = \sum_{n=0}^{N_y-1} \mathbf{u}(y_n) e^{- \frac{2\pi i}{N_y} k_y n}.
\end{equation}

\begin{figure}
    \centering
    \includegraphics[width=\columnwidth]{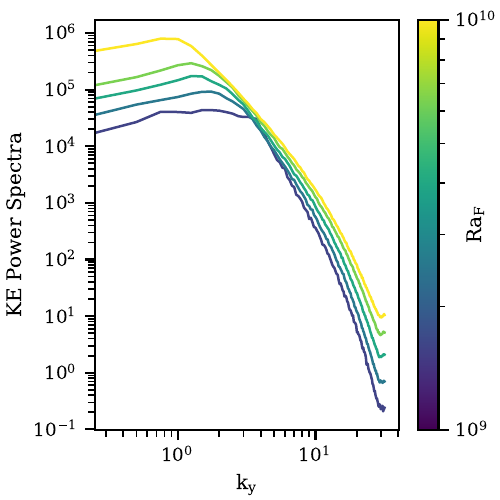}
    \caption{The kinetic energy power spectra for selected representative free-slip, $\text{Ta}=10^8$ cases.}
    \label{fig:Spectra}
\end{figure}

Then we compute the power spectral density as the product of the velocity Fourier transform and its complex conjugate $\mathbf{U*}$, normalised by the horizontal resolution squared, $\mathcal{P} = \frac{1}{N_y^2} \mathbf{U}(k_y)\mathbf{*} \cdot \mathbf{U}(k_y)$. Representative power spectra for free-slip cases at $\text{Ta}=10^8$ are shown in \autoref{fig:Spectra}; we have selected cases that have $\Rf > 10^9$ as they are highly supercritical and convectively dominated. The spectra displays expected features - an energy plateau at low wavenumber, and then an energy cascade as $k$ increases. The power carried at large $k_y$ is orders of magnitudes less than that at low $k_y$, and there is no significant `pile-up' of power at large $k_y$, suggesting that even the modest resolutions used here are sufficient to resolve the small-scale dynamics.

\subsection{Heat Transport Scalings}
\begin{figure*}
    \centering
    \includegraphics[width=\textwidth]{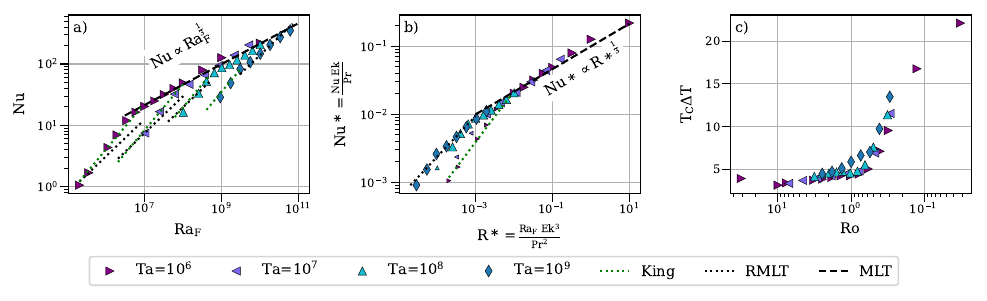}
    \caption{\textbf{a)} Nu vs $\Rf$ for the free-slip cases. The MLT scaling ($\text{Nu} \propto {\Rf}^\frac{1}{3}$) is shown by a black dashed line. The \citet{King12} scaling ($\text{Nu} \propto {\Rf}^\frac{3}{4} \text{Ta}^{-\frac{1}{2}}$) is shown with a green dotted line, and the RMLT scaling ($\text{Nu} \propto \Rf^\frac{3}{5}\text{Ta}^{-\frac{2}{5}}$) is shown with a black dotted line. It can be seen that the rotationally-dominated regimes follow the \citet{King12} scalings better than the RMLT scalings, and that the convectively-dominated regime follows the diffusion-free scaling. \textbf{b)} The diffusion-free Nusselt number $\text{Nu}*$ and the diffusion-free flux-based Rayleigh number $R*$ as calculated in \citet{Hadjerci24} for the free-slip cases. Smaller symbols represent cases with a Nusselt number less than $1.7\times10^1$, indicating cases that are closer to convective onset. \textbf{c)} $\Delta T$ across the domain multiplied by the temperature scale $T_\text{C}$, against the Rossby number as defined in \autoref{eq:Rossby}. As the rotational effects increase, the $\Delta T$ across the domain increases. When Ro is low, the $\Delta T$ across the domain is constant.}
    \label{fig:FS_NuRa}
\end{figure*}

\begin{figure*}
    \centering
    \includegraphics[width=\textwidth]{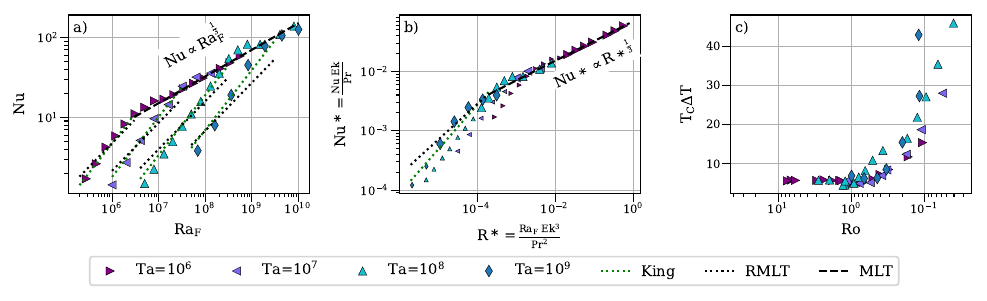}
    \caption{As in \autoref{fig:FS_NuRa}, but for cases with no-slip boundary conditions. a) No-slip Nu vs $\Rf$ for different Taylor number. b) The diffusion-free Nusselt number $\text{Nu}*$ vs the diffusion-free flux-based Rayleigh number $R*$. c) $\Delta T$ across the domain multiplied by the temperature scale $T_\text{C}$, against Rossby number, as defined in \autoref{eq:Rossby}. As above, diffusion-fr  ee scalings are in black and diffusion-limited scalings are in green, with dotted lines denoting rotating scalings and dashed lines giving non-rotating scalings.}
    \label{fig:NS_NuRa}
\end{figure*}

The heat transport scaling with Rayleigh number for free-slip cases are shown in \hyperref[fig:FS_NuRa]{\autoref{fig:FS_NuRa}a}, and the no-slip cases are shown in \hyperref[fig:NS_NuRa]{\autoref{fig:NS_NuRa}a}. For both cases, a transition in behaviour is seen from low Rossby number cases to high Rossby number, with the change in behaviour occuring around $\text{Ro}\simeq1$, similar to that found in \citet{King12}. However, whereas \citet{King12} found that for the high Rossby number cases, the heat transport scaling exhibited the boundary-limited scaling laws, here we see both types of boundary condition display the same behaviour as the findings of \citet{Kazemi22} - that the heat transport scaling for net-zero heated and cooled convection follow the diffusion-free mixing-length scaling of $\text{Nu} \propto {\Rf}^\frac{1}{3}$. 

The scaling behaviour of the low $\text{Ro}$ cases shows more dependence on boundary conditions. For the no-slip boundary conditions, these rotationally-constrained cases follow the scaling laws predicted by \citet{King12} of $\text{Nu} \propto \text{Ta}^{-\frac{2}{5}}{\Rf}^\frac{3}{5}$ quite closely, and do not fit well to the diffusion-free scalings predicted by RMLT \citep[e.g.][]{Julien12, Currie20, Hadjerci24, Lance24}. With free-slip boundaries, the Nu vs $\Rf$ scaling is again similar to the \citet{King12} scaling, although the agreement is less clear compared to the no-slip cases. For the $\text{Ta}=10^9$ cases, there is a possible transition from the boundary-limited scaling to the diffusion-free scaling expected for the low Ro regime. We suspect that this regime is not seen for the lower Taylor numbers due to the relatively low Nusselt numbers explored here, the rotationally-constrained cases at lower Taylor numbers have $\text{Nu}\lesssim30$, and are relatively close to onset, as such, conduction is still playing a role in the overall flux transport. For the $\text{Ta}=10^9$ cases where the RMLT scaling is observed (such as \hyperref[fig:DynamicCompar]{\autoref{fig:DynamicCompar}h}), the rotating regime has $\text{Nu}\gtrsim100$. As the Taylor number increases, so low-Rossby cases can be obtained with a relatively high $\Rf$, we expect the parameter space where RMLT scaling can be observed to become larger (although less computationally accessible). This is consistent with the extreme-parameter simulations of \cite{Song24b}, who observe a transition from boundary-limited to diffusion-free rotating scalings only for values of $\text{Nu}\gtrsim20$. 

As discussed in \autoref{sec:2.1}, we can define diffusivity-free Rayleigh and Nusselt numbers, such that $\Rf* = \Rf\text{Ek}^3/\text{Pr}$ and $\text{Nu}*=\text{NuEk}/\text{Pr}$. In this format, the expected RMLT scaling will be $\text{Nu}* \sim R*^\frac{3}{5}$, and the MLT scaling is $\text{Nu}* \sim R*^\frac{1}{3}$, as reported by \citet{Hadjerci24}. Our data can be seen in this format in both \hyperref[fig:FS_NuRa]{\autoref{fig:FS_NuRa}b} and \hyperref[fig:NS_NuRa]{\autoref{fig:NS_NuRa}b}. In this format, if a scaling is diffusion-free, then we expect the data to collapse down onto a master curve (and by extension, deviation from the master curve suggests the dynamics are not diffusion-free). Again, the diffusion-free behaviour of the convectively-dominated regime is clear, but the scaling of the rotationally-constrained regimes is less conclusive. The free-slip cases seem to show that the higher Taylor number cases follow the RMLT scaling, whereas the lower $\text{Ta}$ cases show the \citet{King12} scaling. Here, we use a smaller symbol size to illustrate cases with $\text{Nu}<20$ - these cases have a lower Nusselt number, i.e. they are closer to onset, and as such the reason for the different scaling could be influenced by the low supercriticality. For no-slip boundary conditions, almost all of the rotationally-constrained cases are close to onset, so it is difficult to draw a conclusion about the scaling, but they seem to agree more with the boundary-limited \citet{King12} scaling than with the diffusion-free RMLT scaling.

Finally, in \hyperref[fig:FS_NuRa]{\autoref{fig:FS_NuRa}c} and \hyperref[fig:NS_NuRa]{\autoref{fig:NS_NuRa}c}, we present how the temperature gradient $\Delta T$ across the domain varies with Ro. This $\Delta T$ is calculated by extracting the horizontally averaged temperature at $z=0$, since $T(z=1)$ is fixed at zero by the boundary conditions. We have re-dimensionalised the temperature gradient by multiplying by the temperature scale $T_\text{C} = d^2 Q/\nu$. These figures display that for non-rotationally-constrained flows $\Delta T$ across the domain is approximately constant. This is consistent with diffusion-free transport in our simulations: because these were done at constant $Q$, variations in $\Rf$ arise solely through variations in diffusivities. Hence, if $\Delta T$ does not depend on $\nu$ or $\kappa$, it should approach a constant value when rotation is unimportant. When rotation is important, by contrast, the convective efficiency is suppressed, and a larger $\Delta T$ is required to transport the same amount of flux across the domain. 

\subsection{Velocity Amplitudes}
\label{sec:VelAmp}

\begin{figure}
    \centering
    \includegraphics[width=\columnwidth]{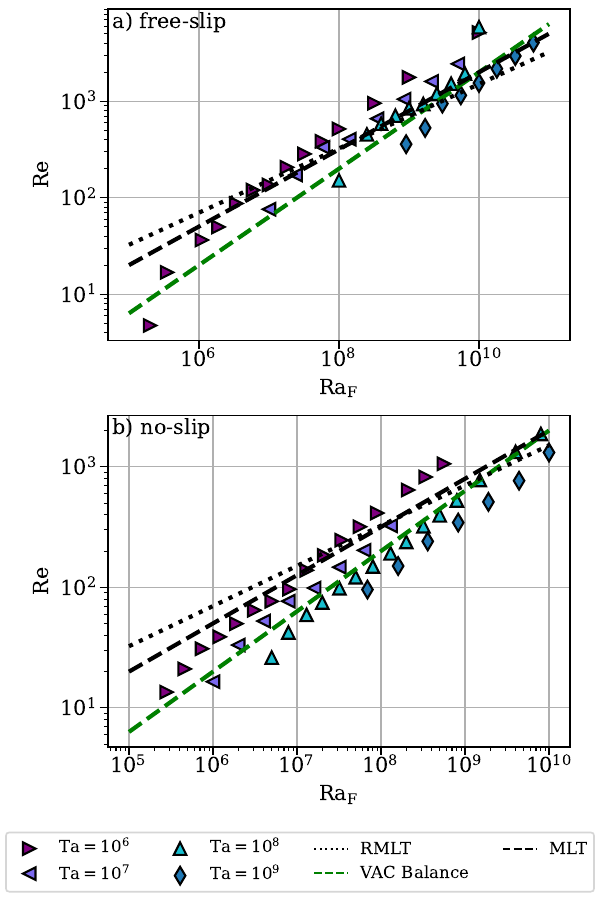}
    \caption{Reynolds number vs $\Rf$ for the free-slip (top) and no-slip (bottom) boundary conditions. The black dashed line represents the MLT scaling, the black dotted line represents the RMLT scaling and the green dashed line represents the diffusion-limited VAC balance scaling. For both boundary conditions, the Reynolds number scaling appears to be better described by the diffusion-limited $\text{Re} \propto {\Rf}^\frac{1}{2}$ scaling, rather than the diffusion-free scalings.}
    \label{fig:Reynolds}
\end{figure}

Flow turbulence can be characterised by the Reynolds number - defined as $\text{Re} = UL/\nu$, where $U$ is the velocity scale, $L$ is the vertical length scale, and a higher Re corresponds to a more turbulent flow. For rapidly rotating diffusion-free flows, the Reynolds number is expected to scale with $\Rf$ like $\text{Re}\propto\Rf^\frac{2}{5} \text{Ta}^{-\frac{1}{10}}$, and in the absence of rotation the diffusion-free scaling is $\text{Re}\propto\Rf^\frac{1}{3}$ \citep{Aurnou20,Bouillaut22,Ecke23}, which is recovered in the 3D simulations and experiments of \citet{Hadjerci24}. For flows where diffusivity is dynamically important it is frequently argued that, in the presence of rotation, a triple balance between viscous, buoyancy (Archimedean) and Coriolis forces (VAC balance) occurs. This can be shown to yield $\text{Re}\propto \Rf^\frac{1}{2} \text{Ta}^{-\frac{1}{6}}$ \citep{aubert_systematic_2001,king_flow_2013,Gastine16}.

The scaling of the Reynolds number with $\Rf$ for our simulations is shown in \autoref{fig:Reynolds}, with the MLT (black dashed line), RMLT (black dotted line) and VAC balance (green dotted line) scalings shown for comparison. The MLT and RMLT predictions do not fit the data, suggesting the velocity amplitude scalings seen here are not diffusion-free. The diffusion-limited VAC balance scaling fits the data reasonably well, despite the majority of the data being in the high-Rossby regime, where the rotation would not be expected to be dynamically important. This figure also shows that increasing the rotation rate decreases the velocity amplitudes; points at similar $\Rf$ have lower values of $\text{Re}$ for higher values of $\text{Ta}$.

\citet{Alexakis06} found that Reynolds number scaling behaves differently for 2D flow to 3D flow, due to differences in energy dissipation (discussed in \autoref{sec:Dissipation}), and similar results were obtained for stratified convection by \citet{Anders17}, finding $\text{Re} \propto \text{Ra}^\frac{3}{4}\ (\propto \Rf^\frac{9}{16})$ for 2D simulations, and $\text{Re} \propto \text{Ra}^\frac{1}{2}$ for 3D $(\propto \Rf^\frac{7}{18})$ and high Ra, supersonic 2D simulations $(\propto \Rf^\frac{5}{12})$. Here, the enhanced turbulence for the 2D flows was attributed to the presence of coherent spinning flows which did not occur in 3D. \citet{Powers23} also found that the presence of long-lived flywheel structures in 2D but not in 3D resulted in an Re scaling that was sensitive to domain dimensions, with $\text{Re} \propto \Rf^{0.45}$ for 2D simulations, and $\text{Re} \propto \Rf^{0.38}$ for 3D simulations, all performed with no-slip boundary conditions and a constant internal heating. We note that for all these cases, 2D simulations display a steeper scaling law than 3D simulations, and so we believe that the 2.5D nature of our simulations is likely to be why our simulations overestimate the expected scaling for the Reynolds number, but display diffusion-free scalings for the Nusselt number.

\subsection{Dissipation}
\label{sec:Dissipation}

\begin{figure}
    \centering
    \includegraphics[width=\columnwidth]{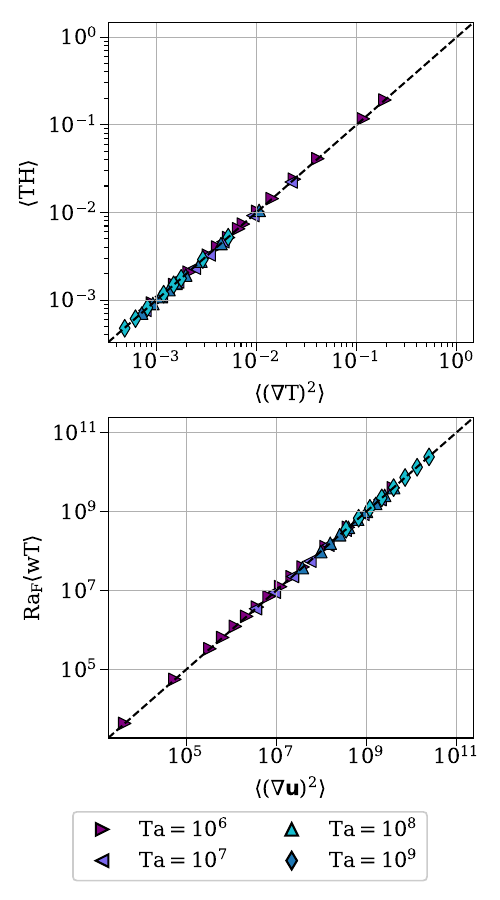}
    \caption{$\langle (\nabla T)^2\rangle$ vs $\langle TH \rangle$ (top); and $\langle (\nabla \mathbf{u})^2 \rangle$ and $\Rf \langle wT \rangle$ (bottom) for free-slip boundary conditions. The black dashed line represents the $y=x$ line, showing that the simulations satisfy the power integrals.}
    \label{fig:ShraimanCompar}
\end{figure}

The efficiency of convection is linked to thermal throttling in the boundary layers \citep{King09, Julien12, Stellmach14}, and the formation and thickness of boundary layers has been found to be linked to thermal and viscous dissipation in a steady-state \citep{Siggia94, Currie17, Long20, Lance24}. More generally, in a convective system the total buoyancy work is balanced by viscous dissipation so constraints on the latter can provide insight into the former.

The exact constraints are encapsulated within the spatio-temporal ``power integrals'', which can be derived from \autoref{eq:MomentumEq} and \autoref{eq:TempEvolution} \citep[see e.g.][]{Shraiman90, Goluskin15}
\begin{equation}
    \langle \varepsilon_u \rangle = \langle (\nabla \mathbf{u} )^2 \rangle = \Rf \langle wT \rangle,    
\end{equation}
where $\varepsilon_u$ is the viscous dissipation; and
\begin{equation}
    \langle \varepsilon_T \rangle = \langle (\nabla T)^2 \rangle = \langle T H(z) \rangle \propto \frac{1}{\text{Nu}}.
\end{equation}
where $\varepsilon_T$ is the thermal dissipation. \autoref{fig:ShraimanCompar} shows these quantities outputted for our simulations, confirming that these relationships hold for all Rossby numbers.

A widely-cited theory developed in \citet{Grossmann00} \citep[see also][]{Grossmann02, Grossmann04, Grossmann11} uses these constraints to derive a theory of convective heat transport. Their theory describes four regimes of non-rotating convection depending on whether each of the thermal and viscous dissipations are bulk- or boundary-dominated; and two sub-regimes for each case depending on whether the thermal boundary layer resides inside the viscous boundary layer, or vice versa. For each regime, they predict different scaling relations for Nu and Re with respect to Ra. According to the Grossmann \& Lohse (GL) theory, the diffusion-free regime (i.e. $\text{Nu}\propto\Rf^\frac{1}{3}$ and $\text{Re}\propto\Rf^\frac{1}{3}$) is only achieved when both thermal and viscous dissipation rates are bulk-dominated. In these regimes, GL theory predicts that the relationship between the dissipation rates and the dynamics should be

\begin{equation}
    \label{eq:GL_Viscous}
    \langle \varepsilon_{\mathbf{u}, \text{bulk}} \rangle \sim \frac{U^3}{L} = \frac{\nu^3}{L^4}\text{Re}^3,
\end{equation}
and
\begin{equation}
    \label{eq:GL_Thermal}
    \langle \varepsilon_{T, \text{bulk}} \rangle \sim \frac{U (\Delta T)^2}{L} = \kappa \frac{(\Delta T)^2}{L^2}\text{Pr}\ \text{Re}.
\end{equation}

First, we investigate where the dissipation is occurring. The GL theory defines a viscous and thermal boundary layer, and then looks at contributions to the total dissipation from within the boundary layers and the bulk. Defining a viscous boundary layer with free-slip boundaries is difficult, so in our own analysis to be consistent between no-slip and free-slip boundary conditions we instead look at contributions to the total dissipation from the bottom, middle, and top thirds of the domain. The justification for this is that if the dissipation is boundary-dominated, then the contribution to total dissipation in one or both of the edge thirds will dominate. Similarly, if the dissipation contribution in the middle third was higher than in the edge thirds, then the dissipation would have to be bulk-dominated. However, high contributions in the edge thirds do not necessarily imply boundary-dominated dissipation, because the width of the boundary layers is much smaller than a third of the domain. Indeed, with our IH\&C function we would expect the majority of the thermal dissipation to occur in the bottom third, independent of any boundary layer dynamics, as this is where all of the heating occurs.

\begin{figure}
    \centering
    \includegraphics[width=\columnwidth]{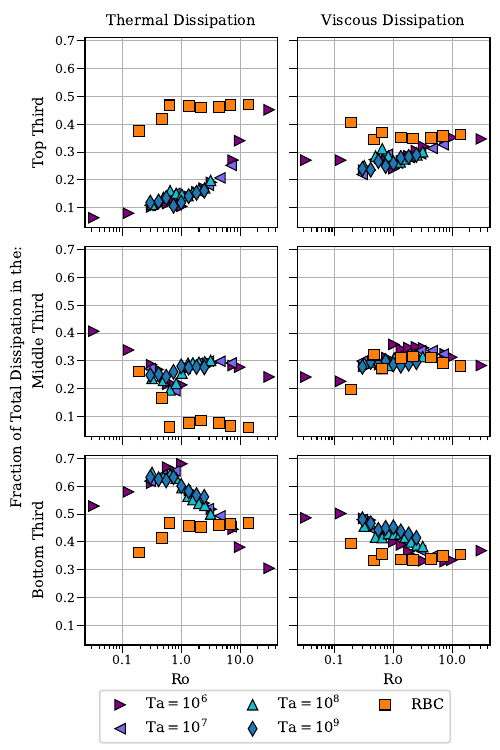}
    \caption{Comparison between the thermal and viscous dissipation contributions in the top, middle and bottom thirds of the domain for internally heated and cooled cases (same symbols as above) compared against Rayleigh-B\'enard convection (RBC) (orange squares). The heated and cooled cases all display a larger proportion of thermal dissipation in the middle third - i.e. in the bulk than the fixed-flux cases, and therefore a lower proportion in the edge thirds (the boundary layers).}
    \label{fig:DissComparison}
\end{figure}

In \autoref{fig:DissComparison}, we compare the thermal and viscous dissipation in each separate third for the free-slip boundary conditions for all Ro, and for comparison, a set of classic free-slip Rayleigh-B\'enard cases driven with a constant fixed flux at the top and bottom boundaries, rather than the internal heating function. The difference in distribution for the thermal dissipation is clear - for the RBC cases, $\langle \varepsilon_T \rangle_H$ is  almost entirely evenly distributed between the top and bottom thirds, with only a small contribution in the middle third (less than ten percent), whereas in the IH\&C cases the middle third contributes over a quarter of the thermal dissipation.  It is also clear that the thermal dissipation in the IH\&C cases display an asymmetry between the top and bottom third, that is not observed in the RBC simulations. This is expected, as the heating is asymmetrical - it is concentrated in the lower third. For the viscous dissipation the difference between fixed-flux and IH\&C behaviour is harder to pinpoint: the apparent low-Rossby deviation in the boundary layers for IH\&C cases (enhanced viscous dissipation in the lower third, and suppression in the upper third) is likely due to the asymmetry in the heating function used. We do not see evidence that viscous dissipation in the bulk is enhanced by internal heating and cooling. 

\begin{figure}
    \centering
    \includegraphics[width=\columnwidth]{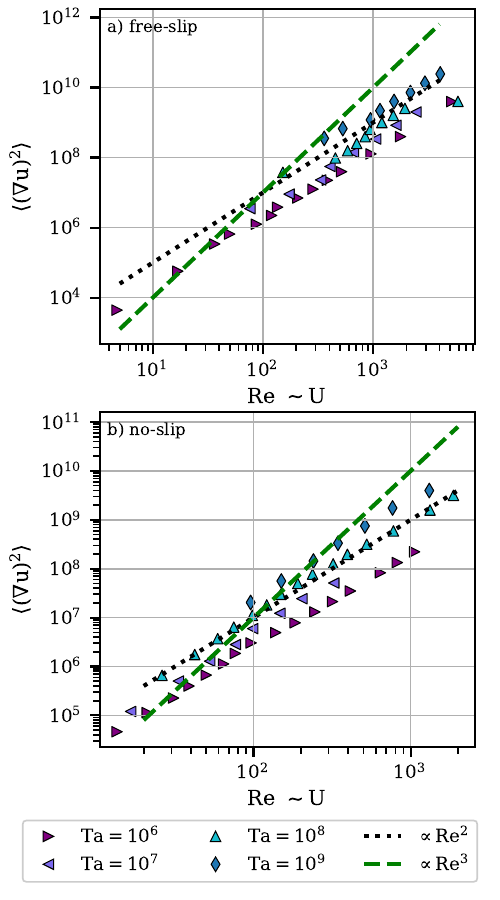}
    \caption{The viscous dissipation rate against the Reynolds number for a) free-slip boundary conditions and b) no-slip boundary conditions. The GL prediction from \autoref{eq:GL_Viscous} is shown by the dashed green line, and the black dashed line shows our best fit of $\langle \varepsilon_\mathbf{u} \rangle \propto \text{Re}^2$.}
    \label{fig:ViscDiss_Re}
\end{figure}

\begin{figure}
    \centering
    \includegraphics[width=\columnwidth]{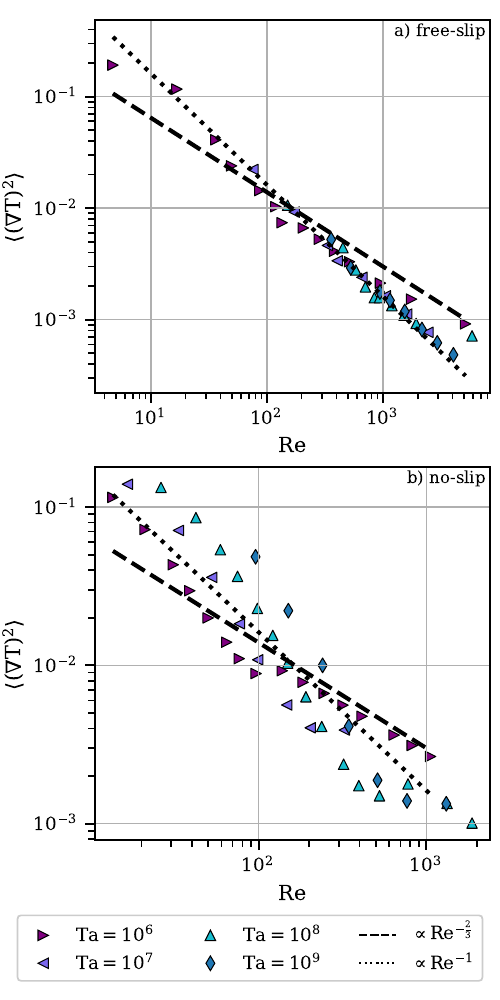}
    \caption{The thermal dissipation rate against the Reynolds number for a) the free-slip boundary conditions and b) the no-slip boundary conditions. The black dotted line shows the $\langle \varepsilon_T\rangle \propto \text{Re}^{-1}$ scaling which fits better for low Re (low Ro) cases. The black dashed line shows the $\langle \varepsilon_T \rangle \propto \text{Re}^{-\frac{2}{3}}$ scaling which is possibly matched by the high $\Rf$, $\text{Ta}=10^6$ cases. }
    \label{fig:ThermDiss_Re}
\end{figure}

We also test the GL predictions from \hyperref[eq:GL_Viscous]{Equations \ref{eq:GL_Viscous} \& \ref{eq:GL_Thermal}}. \autoref{fig:ViscDiss_Re} gives the viscous dissipation against the Reynolds number. We see that the GL prediction is not recovered by the data, and instead a scaling of $\langle \varepsilon_\mathbf{u} \rangle \propto \text{Re}^2$ is observed. We suspect that, similarly to the discussion in \autoref{sec:VelAmp}, this could be linked to the 2.5D nature of the simulations.
\autoref{fig:ThermDiss_Re} shows how the thermal dissipation rate varies with Re, and demonstrates that the GL prediction of $\langle \varepsilon_T \rangle \propto \text{Re}$ is also not obtained; instead we see a scaling of $\langle \varepsilon_T \rangle \propto \text{Re}^{-1}$ for low-Ro cases, and potentially a transition to $\propto \text{Re}^{-\frac{2}{3}}$ for less rotationally constrained, higher Re cases.

Our results suggest that although the heat transport in our simulations follows diffusion-free scalings, the manner in which this is achieved is different than in the GL theory. Prior work, such as \citet{Wang21}, which extends GL arguments to non-rotating IH convection, likewise found some discrepancies with GL theory: for example they observed viscous dissipation scaling roughly with $\text{Re}^2$ in keeping with our findings here. However, they argued that these departures from the theory were a consequence of the parameter space explored and resulting boundary layer dominance. The fact that we observe similar discrepancies in our system, despite the enhanced bulk thermal dissipation and diffusion-free $\text{Nu}(\Rf)$ relation (which suggests that the bulk dynamics are important and the boundary layers do not dominate), suggests that they may reflect more fundamental disagreements between 2D convection and the theory. We suspect some of these discrepancies arise due to the difference in 2D and 3D flows, as noted above.

\section{Discussion and Conclusions}
\label{sec:Discussion}
We have presented an investigation into the dynamics of internally-heated rotating convection with both no-slip and free-slip boundary conditions in 2.5D. We have shown that in these set-ups, for high Rossby numbers, the convective heat transport appears to be independent of diffusivity. Hence, these models can be used as a test-bed for investigating diffusion-free convection in a computationally accessible parameter space.
We have found that for the rotationally-constrained regimes, the heat transport is harder to classify, but there is some evidence that at the highest Taylor numbers it approaches diffusion-free scalings, especially for free-slip boundary conditions. This mirrors the no-slip experimental findings of \citet{Lepot18} and numerical simulations by \citet{Hadjerci24}, both of which report diffusion-free heat transport when radiatively driving convection.

In contrast, the velocity amplitudes are observed to display diffusion-limited scalings, rather than the diffusion-free behaviour observed in the 3D simulations of \citet{Hadjerci24}. We believe this discrepency is caused by the 2.5D nature of the simulations, supported by prior work \citep[e.g.][]{Alexakis06, Anders17, Powers23} also reporting steeper velocity amplitude scalings for 2D simulations than 3D simulations of the same set-ups. Motivated by the exact link between convective dynamics and dissipation, we investigated the thermal and viscous dissipation in our simulations, and compared this to the dissipation in boundary-driven convection. We find that the presence of an internal heating function is associated with increased thermal dissipation in the bulk, reducing the importance of the thermal boundary layer. This in turn allows the efficiency of the convection to increase, which contributes to obtaining the diffusion-free scaling. 
Finally, we examined the applicability of the Grossmann and Lohse theory of convective heat transport \citep{Grossmann00} to this set-up, by testing its predictions of the thermal and viscous dissipation scale with the Reynolds number. We find that these predictions differed appreciably to the scalings examined in the classical GL theory, suggesting that the route by which our results obtain diffusion-free heat transport is likewise different.

In future work, we aim to clarify which aspects of our results depend critically on whether the flow is in 2D or 3D. For example, it may well be that the velocity amplitude and dissipation rates scale differently for 3D flows, even if the heat transport does not. We also aim to investigate the sensitivity of the dynamics to the shape and spatial extent of the heating and cooling functions. Other aspects, such as the dependence of the transport on the Prandtl  number in this set-up also await future investigation.

Nonetheless, our results offer promising news for the study of astrophysical convection. Real stars have vanishingly low microphysical diffusivities, so a long-help theoretical expectation is that they are likely to exhibit diffusion-free convection \citep[e.g.][]{Spiegel71}, as encapsulated in mixing-length theory. Yet the  convection will also be influenced by stratification, magnetism and rotation, none of which are present in classical MLT. To understand the effects of these `complicating factors' we must first understand how diffusion-free convection behaves in their absence, which has been challenging to achieve in classical Rayleigh-B\'enard-like set-ups. It is striking that even in our 2.5D simulations the internally heated and cooled set-ups achieve mixing-length heat transport scalings, providing a computationally inexpensive way to study diffusion-free convection with or without rotation. 

We believe this set-up may be a promising test-bed for incorporating other physical effects such as magnetism. This is because the diffusion-free nature of the transport here will allow us to delineate these physical effects from ones that arise due to numerical limitations, such as boundary layers of  finite width. In this way, a thorough understanding of how `astrophysical' convection behaves may be obtainable in an achievable parameter space, and extrapolated out to unreachable, astrophysical values.

\section*{Acknowledgements}
TJH was supported by a Science and Technology Facilities Council (STFC) PhD studentship (ST/W507453/1). MB and NT gratefully acknowledge support from the STFC (ST/Y002156/1). LC gratefully acknowledges support from the STFC (ST/X001083/1). The authors would like to thank Evan Anders and Whitney Powers (UC Boulder) for insightful discussions which contributed to this paper. We also thank the referee for a thoughtful review which improved this manuscript. Part of this research was performed while the author was visiting the Institute for Pure and Applied Mathematics (IPAM), which is supported by the National Science Foundation (Grant No. DMS-1925919). We also thank the Isaac Newton Institute for Mathematical Sciences, Cambridge, for support and hospitality during the programme “Frontiers in Dynamo Theory: From the Earth to the Stars" (DYT2) where work on this paper was undertaken; this was supported by EPSRC grant no EP/R014604/1. This paper draws on simulations that were carried out on the University of Exeter supercomputer Isca, and on the DiRAC Data Intensive service at Leicester (DiaL), operated by the University of Leicester IT Services, which forms part of the STFC DiRAC HPC Facility (\url{www.dirac.ac.uk}). The DiaL equipment was funded by BEIS capital funding via STFC capital grants ST/K000373/1 and ST/R002363/1 and STFC DiRAC Operations grant ST/R001014/1. DiRAC is part of the National e-Infrastructure.

\section*{Data Availability}

The codes used to produce the simulations in this paper and selected outputs from the simulations have been uploaded to Zenodo and are available for download there (\url{https://doi.org/10.5281/zenodo.15837412}).



\bibliographystyle{mnras}
\bibliography{Bibliography} 



\appendix

\section{Table of Simulations}
\label{app:SimParams}
A summary of the input parameters, and some outputted quantities, for the simulations is given in \autoref{tab:SimParams}.

\begin{table*}
\centering
\caption{A summary of input and output quantities for the simulations described in the text. Column 1-4 are the input parameters; flux-based Rayleigh number $\Rf$, Taylor number $\text{Ta}$, convective flux-based Rossby number $\text{Ro}_\text{c, F}$ and the horizontal boundary condition (either free- or no-slip). Column 5 gives the resolution of each simulation ($Ny \times Nz$), which is the number of Fourier (Chebyshev) modes in the horizontal (vertical) directions after de-aliasing has been applied (i.e. the total number of modes in each direction is 3/2 times the numbers quoted here. Column 6 gives the total sim time that the case was run for, in units of viscous time $\tau_\nu$. Column 7 gives the percentage discrepancy between the imposed flux and the total flux in the simulation, used to check whether time-averaging was sufficient. The final five columns give some output values derived from the simulation results, the Nusselt number Nu calculated as $\text{Nu} = 1 + \langle wT \rangle / \langle -dT/dz \rangle$, the Reynolds number Re calculated as $\text{Re}=\langle \sqrt{\mathbf{u}\cdot\mathbf{u}}\rangle$, the Rossby number Ro calculated as $\langle |\mathbf{\nabla \times u}| / 2|\mathbf{\Omega}|\rangle_V$and the viscous and thermal dissipations, $\langle \varepsilon_u \rangle$ and $\langle \varepsilon_T \rangle$, calculated as described in \autoref{sec:Dissipation}. All cases are run with aspect ratio $\Gamma=4$, $\text{Pr}=1$ and at $\theta=5^\circ$.}
\label{tab:SimParams}
\begin{tabular}{@{}cccccccccccc@{}}
\toprule
$\Rf$ & Ta & Ro$_\text{c, F}$ & \begin{tabular}[c]{@{}c@{}}Boundary\\ Condition\end{tabular} & Resolution & \begin{tabular}[c]{@{}c@{}}Run time\\ ($\tau_\nu$)\end{tabular} & \begin{tabular}[c]{@{}c@{}}$\frac{\Delta F}{F_\text{tot}}$\\ (\%)\end{tabular} & Nu & Re & Ro & $\langle \varepsilon_u \rangle$ & $\langle \varepsilon_T \rangle$ \\ \midrule
2.0e+05 & 1.0e+06 & 0.45 & free-slip & 128$\times$64 & 5.0 & 0.42 & 1.06 & 4.73 & 0.03 & 4.4e+03 & 1.9e-01 \\
3.5e+05 & 1.0e+06 & 0.59 & free-slip & 128$\times$64 & 5.0 & 0.42 & 1.68 & 16.83 & 0.13 & 5.7e+04 & 1.2e-01 \\
1.1e+06 & 1.0e+06 & 1.05 & free-slip & 128$\times$64 & 5.0 & 0.43 & 4.35 & 36.43 & 0.31 & 3.4e+05 & 4.1e-02 \\
1.9e+06 & 1.0e+06 & 1.38 & free-slip & 128$\times$64 & 5.0 & 0.42 & 6.97 & 49.77 & 0.40 & 6.5e+05 & 2.4e-02 \\
3.4e+06 & 1.0e+06 & 1.84 & free-slip & 128$\times$64 & 5.0 & 0.44 & 11.94 & 87.13 & 0.58 & 1.2e+06 & 1.4e-02 \\
5.9e+06 & 1.0e+06 & 2.43 & free-slip & 128$\times$64 & 5.0 & 1.06 & 16.44 & 119.86 & 0.77 & 2.2e+06 & 1.0e-02 \\
1.0e+07 & 1.0e+06 & 3.16 & free-slip & 128$\times$64 & 5.0 & 0.59 & 20.38 & 135.75 & 1.02 & 3.8e+06 & 7.4e-03 \\
1.8e+07 & 1.0e+06 & 4.24 & free-slip & 128$\times$64 & 5.0 & 1.67 & 25.00 & 206.60 & 1.37 & 7.0e+06 & 6.6e-03 \\
3.2e+07 & 1.0e+06 & 5.66 & free-slip & 128$\times$64 & 5.0 & 2.00 & 31.85 & 284.34 & 1.83 & 1.2e+07 & 5.3e-03 \\
5.7e+07 & 1.0e+06 & 7.55 & free-slip & 128$\times$64 & 5.0 & 1.51 & 39.95 & 384.07 & 2.45 & 2.2e+07 & 4.1e-03 \\
1.0e+08 & 1.0e+06 & 10.00 & free-slip & 128$\times$64 & 5.0 & 1.53 & 49.09 & 516.90 & 3.25 & 3.9e+07 & 3.3e-03 \\
3.2e+08 & 1.0e+06 & 17.89 & free-slip & 128$\times$128 & 2.5 & 1.88 & 79.96 & 955.00 & 7.58 & 1.3e+08 & 2.1e-03 \\
1.0e+09 & 1.0e+06 & 31.62 & free-slip & 128$\times$128 & 2.5 & 2.60 & 126.86 & 1771.76 & 9.85 & 4.0e+08 & 1.5e-03 \\
1.0e+10 & 1.0e+06 & 100.00 & free-slip & 128$\times$128 & 2.5 & 1.21 & 219.04 & 5166.95 & 31.22 & 3.9e+09 & 9.2e-04 \\
1.0e+07 & 1.0e+07 & 1.00 & free-slip & 128$\times$128 & 5.0 & 0.12 & 7.47 & 75.96 & 0.29 & 3.5e+06 & 2.2e-02 \\
2.4e+07 & 1.0e+07 & 1.55 & free-slip & 128$\times$128 & 5.0 & 0.12 & 16.71 & 172.28 & 0.48 & 9.0e+06 & 9.2e-03 \\
5.9e+07 & 1.0e+07 & 2.43 & free-slip & 128$\times$128 & 5.0 & 1.53 & 32.23 & 332.98 & 0.78 & 2.3e+07 & 4.6e-03 \\
1.4e+08 & 1.0e+07 & 3.74 & free-slip & 128$\times$128 & 5.0 & 1.34 & 46.67 & 404.70 & 1.18 & 5.5e+07 & 3.4e-03 \\
3.5e+08 & 1.0e+07 & 5.92 & free-slip & 128$\times$128 & 5.0 & 1.50 & 66.72 & 660.44 & 1.87 & 1.4e+08 & 2.4e-03 \\
8.5e+08 & 1.0e+07 & 9.22 & free-slip & 128$\times$128 & 5.0 & 0.90 & 94.78 & 1052.52 & 2.90 & 3.4e+08 & 1.6e-03 \\
2.1e+09 & 1.0e+07 & 14.49 & free-slip & 128$\times$128 & 5.0 & 1.27 & 139.37 & 1608.69 & 4.56 & 8.3e+08 & 1.1e-03 \\
5.0e+09 & 1.0e+07 & 22.36 & free-slip & 128$\times$128 & 7.1 & 0.92 & 204.47 & 2446.07 & 7.04 & 2.0e+09 & 7.7e-04 \\
1.0e+08 & 1.0e+08 & 1.00 & free-slip & 128$\times$128 & 5.0 & 0.90 & 16.35 & 151.32 & 0.32 & 3.8e+07 & 1.1e-02 \\
2.5e+08 & 1.0e+08 & 1.58 & free-slip & 256$\times$64 & 5.0 & 1.14 & 33.28 & 453.90 & 0.50 & 9.7e+07 & 4.4e-03 \\
4.0e+08 & 1.0e+08 & 2.00 & free-slip & 256$\times$64 & 5.0 & 1.11 & 52.48 & 586.03 & 0.64 & 1.6e+08 & 2.8e-03 \\
6.4e+08 & 1.0e+08 & 2.53 & free-slip & 256$\times$64 & 5.0 & 1.51 & 72.18 & 708.90 & 0.82 & 2.5e+08 & 2.0e-03 \\
1.0e+09 & 1.0e+08 & 3.16 & free-slip & 256$\times$64 & 5.0 & 1.05 & 87.35 & 846.44 & 1.03 & 4.0e+08 & 1.6e-03 \\
1.6e+09 & 1.0e+08 & 4.00 & free-slip & 256$\times$64 & 5.0 & 1.39 & 97.89 & 938.18 & 1.30 & 6.3e+08 & 1.6e-03 \\
2.5e+09 & 1.0e+08 & 5.00 & free-slip & 256$\times$64 & 5.0 & 0.88 & 116.25 & 1198.43 & 1.58 & 1.0e+09 & 1.3e-03 \\
4.0e+09 & 1.0e+08 & 6.32 & free-slip & 256$\times$64 & 5.0 & 0.68 & 139.74 & 1523.07 & 2.00 & 1.6e+09 & 1.1e-03 \\
6.3e+09 & 1.0e+08 & 7.94 & free-slip & 256$\times$64 & 5.0 & 1.08 & 165.47 & 1940.55 & 2.51 & 2.5e+09 & 9.2e-04 \\
1.0e+10 & 1.0e+08 & 10.00 & free-slip & 256$\times$64 & 5.0 & 1.35 & 207.29 & 5881.89 & 3.16 & 4.0e+09 & 7.1e-04 \\
9.1e+08 & 1.0e+09 & 0.95 & free-slip & 128$\times$128 & 2.5 & 0.29 & 28.78 & 359.17 & 0.30 & 3.5e+08 & 5.2e-03 \\
1.7e+09 & 1.0e+09 & 1.30 & free-slip & 128$\times$128 & 2.5 & 0.43 & 48.89 & 529.75 & 0.41 & 6.7e+08 & 2.9e-03 \\
3.0e+09 & 1.0e+09 & 1.73 & free-slip & 128$\times$128 & 2.5 & 0.75 & 82.63 & 945.95 & 0.55 & 1.2e+09 & 1.8e-03 \\
5.5e+09 & 1.0e+09 & 2.35 & free-slip & 128$\times$128 & 2.5 & 0.27 & 107.06 & 1153.39 & 0.74 & 2.2e+09 & 1.5e-03 \\
1.0e+10 & 1.0e+09 & 3.16 & free-slip & 128$\times$128 & 2.5 & 0.73 & 147.10 & 1547.33 & 1.00 & 4.0e+09 & 1.2e-03 \\
1.8e+10 & 1.0e+09 & 4.24 & free-slip & 128$\times$128 & 2.5 & 0.74 & 204.76 & 2176.69 & 1.34 & 7.2e+09 & 8.2e-04 \\
3.3e+10 & 1.0e+09 & 5.74 & free-slip & 128$\times$128 & 2.5 & 0.78 & 274.11 & 2950.12 & 1.82 & 1.3e+10 & 6.2e-04 \\
6.0e+10 & 1.0e+09 & 7.75 & free-slip & 128$\times$128 & 2.5 & 0.54 & 351.99 & 4062.46 & 2.46 & 2.4e+10 & 4.8e-04 \\
\hline
\end{tabular}
\end{table*}
\begin{table*}
\contcaption{}
\begin{tabular}{cccccccccccc}
\toprule
$\Rf$ & Ta & Ro$_\text{c, F}$ & \begin{tabular}[c]{@{}c@{}}Boundary\\ Condition\end{tabular} & Resolution & \begin{tabular}[c]{@{}c@{}}Run time\\ ($\tau_\nu$)\end{tabular} & \begin{tabular}[c]{@{}c@{}}$\frac{\Delta F}{F_\text{tot}}$\\ (\%)\end{tabular} & Nu & Re & Ro & $\langle \varepsilon_u \rangle$ & $\langle \varepsilon_T \rangle$ \\ \midrule
2.8e+05 & 1.0e+06 & 0.53 & no-slip & 128$\times$64 & 5.0 & 0.53 & 1.70 & 13.54 & 0.11 & 4.6e+04 & 1.2e-01 \\
4.6e+05 & 1.0e+06 & 0.68 & no-slip & 128$\times$64 & 5.0 & 0.26 & 2.63 & 21.06 & 0.17 & 1.1e+05 & 7.2e-02 \\
7.4e+05 & 1.0e+06 & 0.86 & no-slip & 128$\times$64 & 5.0 & 0.17 & 4.25 & 30.97 & 0.31 & 2.3e+05 & 4.3e-02 \\
1.2e+06 & 1.0e+06 & 1.10 & no-slip & 128$\times$64 & 5.0 & 0.34 & 5.87 & 38.86 & 0.41 & 4.0e+05 & 3.0e-02 \\
1.9e+06 & 1.0e+06 & 1.38 & no-slip & 128$\times$64 & 5.0 & 0.37 & 8.26 & 50.03 & 0.51 & 6.7e+05 & 2.0e-02 \\
3.1e+06 & 1.0e+06 & 1.76 & no-slip & 128$\times$64 & 5.0 & 0.40 & 11.11 & 64.54 & 0.69 & 1.1e+06 & 1.4e-02 \\
5.0e+06 & 1.0e+06 & 2.24 & no-slip & 128$\times$64 & 5.0 & 0.41 & 13.37 & 76.64 & 0.74 & 1.9e+06 & 1.1e-02 \\
8.1e+06 & 1.0e+06 & 2.85 & no-slip & 128$\times$64 & 5.0 & 0.40 & 16.06 & 96.56 & 0.96 & 3.0e+06 & 8.9e-03 \\
1.3e+07 & 1.0e+06 & 3.61 & no-slip & 128$\times$64 & 5.0 & 1.23 & 16.82 & 138.49 & 1.37 & 4.9e+06 & 9.2e-03 \\
2.1e+07 & 1.0e+06 & 4.58 & no-slip & 128$\times$64 & 5.0 & 1.41 & 19.21 & 184.09 & 1.44 & 7.8e+06 & 7.8e-03 \\
3.4e+07 & 1.0e+06 & 5.83 & no-slip & 128$\times$64 & 5.0 & 1.22 & 22.71 & 246.04 & 1.81 & 1.3e+07 & 6.7e-03 \\
5.6e+07 & 1.0e+06 & 7.48 & no-slip & 128$\times$64 & 4.4 & 1.97 & 26.66 & 318.28 & 2.37 & 2.1e+07 & 5.6e-03 \\
9.0e+07 & 1.0e+06 & 9.49 & no-slip & 128$\times$64 & 5.0 & 1.06 & 31.02 & 413.73 & 2.96 & 3.5e+07 & 4.8e-03 \\
2.1e+08 & 1.0e+06 & 14.49 & no-slip & 128$\times$128 & 2.5 & 1.78 & 40.78 & 642.29 & 6.63 & 8.2e+07 & 3.6e-03 \\
3.4e+08 & 1.0e+06 & 18.44 & no-slip & 128$\times$128 & 2.5 & 1.44 & 48.20 & 823.14 & 5.77 & 1.3e+08 & 3.1e-03 \\
5.6e+08 & 1.0e+06 & 23.66 & no-slip & 128$\times$128 & 2.5 & 1.41 & 58.35 & 1059.25 & 7.43 & 2.2e+08 & 2.7e-03 \\
1.0e+06 & 1.0e+07 & 0.32 & no-slip & 128$\times$128 & 5.0 & 0.47 & 1.43 & 16.51 & 0.06 & 1.2e+05 & 1.4e-01 \\
2.0e+06 & 1.0e+07 & 0.45 & no-slip & 128$\times$128 & 5.0 & 1.98 & 2.72 & 33.24 & 0.11 & 5.0e+05 & 7.1e-02 \\
4.0e+06 & 1.0e+07 & 0.63 & no-slip & 128$\times$128 & 5.0 & 1.19 & 5.14 & 52.55 & 0.18 & 1.3e+06 & 3.6e-02 \\
7.9e+06 & 1.0e+07 & 0.89 & no-slip & 128$\times$128 & 5.0 & 1.58 & 9.61 & 76.78 & 0.33 & 2.8e+06 & 1.8e-02 \\
1.6e+07 & 1.0e+07 & 1.26 & no-slip & 128$\times$256 & 5.0 & 0.03 & 14.45 & 98.74 & 0.39 & 6.0e+06 & 1.1e-02 \\
3.2e+07 & 1.0e+07 & 1.79 & no-slip & 128$\times$128 & 5.0 & 0.22 & 24.51 & 147.05 & 0.55 & 1.2e+07 & 5.6e-03 \\
6.3e+07 & 1.0e+07 & 2.51 & no-slip & 128$\times$128 & 5.0 & 0.38 & 31.85 & 203.03 & 0.78 & 2.4e+07 & 4.0e-03 \\
1.3e+08 & 1.0e+07 & 3.61 & no-slip & 128$\times$128 & 5.0 & 0.88 & 35.69 & 324.23 & 1.15 & 5.1e+07 & 3.9e-03 \\
5.0e+06 & 1.0e+08 & 0.22 & no-slip & 128$\times$128 & 5.0 & 0.37 & 1.49 & 26.18 & 0.04 & 6.6e+05 & 1.3e-01 \\
7.9e+06 & 1.0e+08 & 0.28 & no-slip & 128$\times$64 & 5.0 & 0.18 & 2.25 & 42.26 & 0.07 & 1.8e+06 & 8.6e-02 \\
1.3e+07 & 1.0e+08 & 0.36 & no-slip & 128$\times$64 & 5.0 & 0.32 & 3.48 & 59.14 & 0.10 & 3.7e+06 & 5.4e-02 \\
2.0e+07 & 1.0e+08 & 0.45 & no-slip & 128$\times$64 & 5.0 & 0.40 & 4.98 & 74.86 & 0.13 & 6.4e+06 & 3.6e-02 \\
3.2e+07 & 1.0e+08 & 0.57 & no-slip & 128$\times$64 & 5.0 & 0.40 & 7.72 & 98.31 & 0.17 & 1.1e+07 & 2.3e-02 \\
5.0e+07 & 1.0e+08 & 0.71 & no-slip & 256$\times$128 & 5.0 & 0.12 & 11.03 & 121.39 & 0.37 & 1.8e+07 & 1.5e-02 \\
8.0e+07 & 1.0e+08 & 0.89 & no-slip & 128$\times$64 & 5.0 & 0.42 & 15.83 & 148.98 & 0.51 & 3.0e+07 & 1.0e-02 \\
1.3e+08 & 1.0e+08 & 1.14 & no-slip & 128$\times$64 & 5.0 & 0.42 & 24.57 & 191.38 & 0.65 & 5.0e+07 & 6.3e-03 \\
2.0e+08 & 1.0e+08 & 1.41 & no-slip & 128$\times$64 & 5.0 & 0.42 & 35.51 & 237.67 & 0.80 & 7.8e+07 & 4.1e-03 \\
3.2e+08 & 1.0e+08 & 1.79 & no-slip & 128$\times$64 & 5.0 & 0.42 & 54.34 & 321.13 & 0.92 & 1.3e+08 & 2.4e-03 \\
5.0e+08 & 1.0e+08 & 2.24 & no-slip & 128$\times$64 & 5.0 & 0.42 & 69.81 & 395.80 & 1.22 & 2.0e+08 & 1.7e-03 \\
8.0e+08 & 1.0e+08 & 2.83 & no-slip & 128$\times$64 & 5.0 & 0.68 & 81.84 & 525.37 & 1.30 & 3.2e+08 & 1.5e-03 \\
1.5e+09 & 1.0e+08 & 3.87 & no-slip & 128$\times$64 & 7.5 & 0.56 & 81.98 & 776.79 & 1.22 & 5.9e+08 & 1.8e-03 \\
4.0e+09 & 1.0e+08 & 6.32 & no-slip & 128$\times$64 & 7.5 & 0.57 & 108.98 & 1328.59 & 1.99 & 1.6e+09 & 1.3e-03 \\
8.0e+09 & 1.0e+08 & 8.94 & no-slip & 128$\times$64 & 7.5 & 0.66 & 138.26 & 1871.62 & 2.81 & 3.2e+09 & 1.0e-03 \\
6.9e+07 & 1.0e+09 & 0.26 & no-slip & 128$\times$128 & 2.5 & 0.13 & 3.83 & 95.88 & 0.12 & 2.0e+07 & 4.8e-02 \\
1.6e+08 & 1.0e+09 & 0.40 & no-slip & 128$\times$256 & 4.2 & 0.03 & 7.97 & 150.30 & 0.12 & 5.6e+07 & 2.2e-02 \\
3.6e+08 & 1.0e+09 & 0.60 & no-slip & 128$\times$256 & 4.2 & 2.49 & 18.97 & 240.61 & 0.20 & 1.4e+08 & 1.0e-02 \\
8.3e+08 & 1.0e+09 & 0.91 & no-slip & 128$\times$256 & 1.9 & 2.43 & 45.02 & 344.80 & 0.33 & 3.3e+08 & 4.1e-03 \\
1.9e+09 & 1.0e+09 & 1.38 & no-slip & 128$\times$128 & 2.5 & 0.12 & 78.07 & 511.62 & 0.44 & 7.5e+08 & 1.9e-03 \\
4.4e+09 & 1.0e+09 & 2.10 & no-slip & 128$\times$128 & 2.5 & 0.43 & 105.66 & 766.88 & 0.67 & 1.7e+09 & 1.4e-03 \\
1.0e+10 & 1.0e+09 & 3.16 & no-slip & 128$\times$128 & 2.5 & 0.42 & 125.36 & 1313.70 & 1.00 & 4.0e+09 & 1.3e-03 \\
\bottomrule
\end{tabular}
\end{table*}

\bsp	
\label{lastpage}
\end{document}